\documentclass{egpubl}

 \JournalSubmission    % uncomment for submission to Computer Graphics Forum
 \electronicVersion % can be used both for the printed and electronic version

\ifpdf \usepackage[pdftex]{graphicx} \pdfcompresslevel=9
\else \usepackage[dvips]{graphicx} \fi

\PrintedOrElectronic

\usepackage{t1enc,dfadobe}

\usepackage{egweblnk}
\usepackage{cite}

\usepackage[cmex10]{amsmath} %for \rm in eq:iso

\title[Topologically Robust 3D Shape Matching via Gradual Deflation and Inflation]%
      {Topologically Robust 3D Shape Matching via\\ Gradual Deflation and Inflation}

\author[A. Genctav \& Y. Sahillioglu \& S. Tari]
       {Asli Genctav\thanks{Corresponding Author (e-mail: asli@ceng.metu.edu.tr)}, Yusuf Sahillioglu,
        and Sibel Tari
        \\
         Middle East Technical University, Computer Engineering Dept., 06800, Ankara, Turkey\\     
       }

\begin{document}

% \teaser{
%  \includegraphics[width=\linewidth]{eg_new}
%  \centering
%   \caption{New EG Logo}
% \label{fig:teaser}
% }

\maketitle

\begin{abstract}

Despite being vastly ignored in the literature, coping with topological noise is an issue of increasing importance, especially as a consequence of the increasing number and diversity of 3D polygonal models that are captured by devices of different qualities or synthesized by algorithms of different stabilities.  One approach for matching 3D shapes under topological noise is to replace the topology-sensitive geodesic distance with distances that are less sensitive to topological changes.  We propose an alternative approach utilising gradual deflation (or inflation) of the shape volume, of which purpose is to bring the pair of shapes to be matched to a \emph{comparable} topology before the search for correspondences. Illustrative experiments using different datasets demonstrate that as the level of topological noise increases, our approach outperforms the other methods in the literature. 
  
%\begin{classification} % according to http://www.acm.org/class/1998/
%\CCScat{Computer Graphics}{I.3.5}{3D Shape Correspondence}{Topological noise}
%\end{classification}

\end{abstract}

%-------------------------------------------------------------------------
\section{Introduction}

Establishing correspondences between a pair of 3D shapes is a crucial problem in computer graphics and vision with many applications including, but not limited to, attribute transfer, morphing, registration, retrieval, statistical modeling, and tracking \cite{Bronstein08, Kaick11}. 3D shape correspondence methods seek a plausible mapping that pair up similar or semantically equivalent surface points of two given shapes. Thanks to the broad application domain of this problem, it comes in various scenarios, such as correspondence between completely or partially similar objects that admit rigid or non-rigid deformations under geometric or topological noise. Recently, we also observe exploitation of context information when a collection of three or more shapes is given to be matched.

In this paper, we address a less-explored scenario that deals with a pair of non-rigid shapes \emph{under topological noise}. Shapes that are extrinsically different by non-rigid deformations, a.k.a isometric shapes, commonly appear in various contexts such as different poses of an articulated object, or two shapes representing different but semantically similar objects (e.g., two different humans). Despite being vastly ignored in the existing literature, topological noise is also a common issue with the increasing number and diversity of 3D polygonal models that are captured by devices of different qualities or synthesized by algorithms of different stabilities. We therefore believe that handling 3D shape correspondence under topological noise makes an important contribution to the computer graphics and vision fields.

There are, to the best of our knowledge, only four methods that address the shape correspondence problem under topological noise, three of which merely replace the topology-sensitive geodesic distance with topologically-robust diffusion distance \cite{Bronstein10, Sharma10, Sharma11}, whereas the other employs a heuristic eigenfunction alignment scheme \cite{Mateus08}.

We propose an alternative approach utilising gradual deflation (or inflation) of the shape volume. The purpose of inflating (or deflating) is to bring the pair of shapes to be matched to a \emph{comparable} topology before the search for correspondences. Roughly speaking, the process of deflation (inflation) is  curvature-dependent erosion (dilation).
The reason that we perform these processes in curvature-sensitive manner is to avoid development of singularities -- cusps and collapsed boundaries. Once the pair of shapes is brought to a topologically comparable form, a 3D isometric shape correspondence algorithm using geodesic distance is employed to find correspondences between topologically comparable forms; the found correspondences between topologically comparable forms are then transferred to the correspondences between the original pair of input shapes.

Though the above-mentioned previous methods handle topology noise to some extent, our approach provides more robust solution as the level of topological noise increases. 

For the purpose of equating the topologies of the pair of shapes, we employ a curvature-sensitive distance field. Specifically, the boundaries of the gradually deflated (or inflated) volumes  are obtained as the level surfaces of the field.
The isometric correspondences are searched between comparable level surfaces sampled from the respective fields computed for each shape of the pair, and then transferred to the original shape surfaces. An immediate advantage of a  field based solution is the replacement of the scale parameter in diffusion distances with a theoretically sound and automatic stopping condition based on Euler's formula. Level surfaces obtained from the field can be analyzed by geodesic distances, which are known to be more accurate than diffusion based counterparts.

Note that, although we primarily target the correspondence problem under topological noise, the proposed scheme can also be used to generate correspondences between shape pairs not suffering from such noise. 

%-------------------------------------------------------------------------
\section{Related Work}
\label{related}

Matching semantically similar shapes is a problem studied in deep depth for various scenarios including completely isometric shape pairs \cite{Ovsjanikov10}\cite{Sahillioglu12b}, partially isometric shape pairs \cite{Sahillioglu13}\cite{Litany16}, non-isometric shape pairs \cite{Panozzo13}\cite{Solomon16}, and a collection of shapes \cite{Cosmo16}\cite{Sahillioglu14}. Compared to these studies, and many other related ones \cite{Kaick11}, matching shapes under noise, especially topological noise, is a problem that is yet at its infancy.

An important set of geometric tools to help develop the existing methods under topological noise category consists of several distance metrics on surface, namely the diffusion metric \cite{Coifman05}, the commute-time metric \cite{Wang11}, and the biharmonic metric \cite{Rustamov10}. These distance metrics are all shape-aware in that they are isometry invariant and insensitive to small topological noises. They achieve this property by relying on a principle based on the spread of heat. Intuitively, these metrics capture the spread of heat over time from a seed surface point when a hot needle is touched into that point. In other words, these multiscale distances measure the amount of heat transferred from the seed to the target in a certain time, where the varying times make the metric multiscale. This measurement is obtained by averaging lengths of multiple paths from the seed to the target point, which is hence more topologically-robust than the common geodesic metric that uses only one path, namely the shortest path. We see \cite{Bronstein10}\cite{Sharma10}\cite{Sharma11} employing these topologically-robust metrics in their respective correspondence methods. Due to the inherent instability problem of these metrics under large topological noise, they are doomed to fail when the noise level is increased.

A different approach to topologically-robust correspondence is the eigenfunction alignment of \cite{Mateus08}. In this study, the unreliable eigenvalues of a large sparse Laplacian are reordered with the aim of handling topological noise. Similar to the other related works, it is able to handle very moderate topological noise.

Yet another approach for the correspondence under topological noise problem uses machine learning in order to deal with this geometrically difficult problem \cite{Wei16}. Main limitation of this method is its restriction to a special class of shapes, e.g., it works on human class only. Obtaining a training set that covers all the plausible poses of a given class is a cumbersome process that is also prone to the possibility of missing some poses out of consideration. Problems in the training will result in inconsistent output correspondences.

A recent deformation-based method strives to alleviate the problem of topological noise in correspondence search \cite{Boscaini14}. In this method, two shapes in question are deformed into a canonical pose that removes all the bending transformations over the surface, hence effectively making the resulting shapes comparable under simpler rigid transformations that do not deal with the complicated bendings \cite{Besl92}. Borrowing ideas from Coulomb energy that models electrostatic repulsion, their model naturally deals with topological noises by tearing up the shape at points of strong repulsion.

There is a very recent work on establishing correspondences between surfaces with different topologies \cite{Park16}. Although the problem addressed in this work is not to deal with topological noise, the ideas can help improve the methods handling noise. Essentially, this work morphs the source shape into a topologically different target shape by using the intermediate implicit surfaces, which is in that sense similar to our work.

Although it is not explored yet, the recent functional correspondence framework for shape correspondence has the potential of handling topological noise thanks to their capability of going beyond point-to-point maps \cite{Ovsjanikov12}\cite{Huang14}. Rather than vertices on shapes, these methods matches real-valued functions over surfaces. They, however, have a disadvantage of converting the optimal functional map into a more user-friendly point-to-point map, an action that is prone to bring artifacts and discontinuities \cite{Ganapathi16}.

\section{The Methods}
\label{sec:method}
%-------------------------------------------------------------------------
%\section{The Method}
%\label{sec:method}

A straightforward means to erode (or dilate) shapes in curvature dependent manner   is to move each point on the shape boundary in the direction of surface normal. If the speed is chosen as proportional to a local feature such as curvature, the erosion (or dilation)  becomes curvature dependent.
Such curvature dependent shape boundary motion can be numerically implemented using the level set method~\cite{osher2006level}.
In this work, instead of directly implementing curvature dependent boundary motion, we use an approximate model from \cite{TSP96} which provides a simpler and more efficient computational model. Furthermore, using the approximate model, we obtain all of the deflated or inflated forms (of varying rate) in a single shot.

The used approximate model yields an approximately curvature-dependent distance field $v$ such that
\begin{equation}
v(\mathbf{x}) \approx \rho \biggl( 1+ \frac{\rho}{2} \; curv(\mathbf{x}) \biggr) \frac{\partial v}{\partial n}(\mathbf{x}) + O\left({\rho^3} \right),
\label{eq:approx_v}
\end{equation}
where $curv(\mathbf{x})$ is the curvature of the level surface of $v$ passing through the point $\mathbf{x}$, $n$ is the direction of the inward (or outward) normal and $\rho$ is a parameter determining the smoothness of the level surfaces.

The field $v$ is the minimizer of
\begin{equation}
\Lambda_{\rho}(v) = \frac{1}{2} \int {\left\{ \rho \, {||\nabla v||}^2 + \frac{\left(1-v\right)^2}{\rho} \right\} d\mathbf{x}},
\label{eq:energy_field}
\end{equation}
subject to the boundary condition $v = 0$. Hence, it can be  obtained by solving the  Euler Lagrange equation of \eqref{eq:energy_field}, which is a linear elliptic PDE. We discretize the PDE on a standard grid via finite-difference method.
%Homogeneous Dirichlet condition is applied on the shape boundary.
In the case of field computation outside the shape volume, \emph{i.e.}, inflation, we impose homogeneous Neumann condition on the grid boundary. We determine the parameter $\rho$ as the maximal radius of the shape volume computed using Euclidean distance transform. The discretization yields a linear system of equations with a sparse and symmetric positive definite system matrix. We solve the resulting system using the Cholesky decomposition based direct solver, specifically the MATLAB built-in CHOLMOD implementation.

We first voxelize the mesh model of a given shape. Once the curvature-sensitive distance field $v$ is computed inside (or outside) the shape volume, we generate a collection of level surfaces giving deflated or inflated  forms of the shape boundary at a variety of rates. We obtain the level surfaces using the MATLAB implementation for extracting iso-surface data from volume data. Note that the level surfaces are closed and connected meshes since the field is a smooth and continuous function. If there are more than one connected component in a level surface, we consider the largest one. The set of field values for which the collection of iso-surfaces are extracted are determined for the levels $t$ from $0$ to $1$ sampled at an interval such as $0.01$ using the formula $v_{max}\,\frac{e^{4t}-1}{e^4-1}$ where $v_{max}$ is the maximum value of the field.

\begin{figure}[htb]
  \centering
  \begin{tabular}{c}
  \includegraphics[width=.26\linewidth]{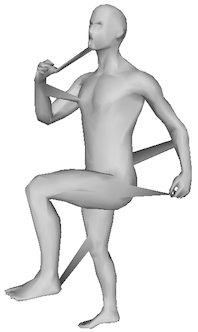}
    \includegraphics[width=.26\linewidth]{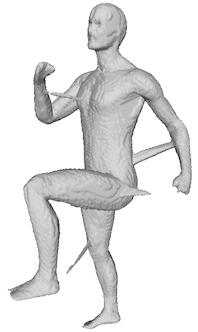}
    \includegraphics[width=.26\linewidth]{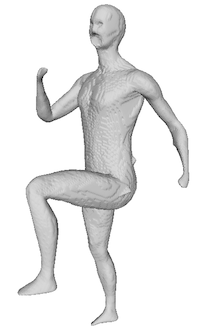}\\
  %(a) \\
  \includegraphics[width=.24\linewidth]{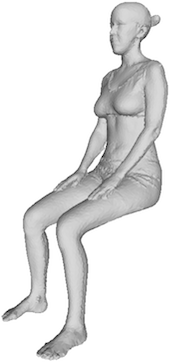}
    \includegraphics[width=.24\linewidth]{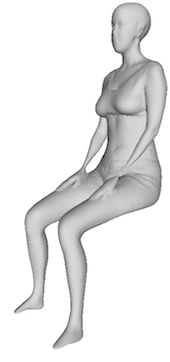}
    \includegraphics[width=.24\linewidth]{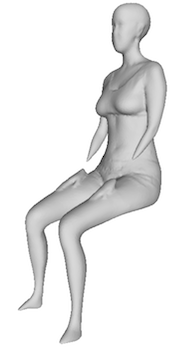}\\
  %(b)\\
  \includegraphics[width=.24\linewidth]{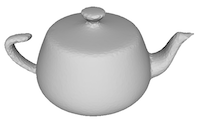}
  \includegraphics[width=.24\linewidth]{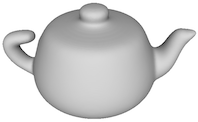}
  \includegraphics[width=.24\linewidth]{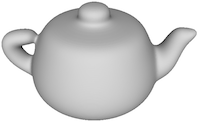}\\
  %(c)\\
  \end{tabular}
  \caption{\label{fig:fieldisosurfaces}
  (Top) A shape with topological noise from SHREC11 robustness benchmark and the inner iso-surfaces of the curvature-sensitive distance field at level $0.1$ and $0.3$. (Middle) A shape from Watertight dataset where the hands are connected to the legs and the inner iso-surfaces of the distance field at level $0.35$ and $0.40$. (Bottom) A teapot shape from Watertight dataset where the handle is cut on the upper joint and the outer iso-surfaces of the distance field at level $0.38$  and $0.43$.}
\end{figure}

In Figure~\ref{fig:fieldisosurfaces}, three illustrative examples are given.  The shape in the top row has topological noise where it is from SHREC11 robustness benchmark~\cite{bronstein2010shrec}. The curvature-sensitive distance field is computed inside the shape volume and the iso-surfaces at level $0.1$ and $0.3$ are the boundaries of the {deflated} forms.  At level $0.3$, the links induced by the topological noise disappear while the main shape structures are preserved. In the middle row, the shape from Watertight dataset~\cite{giorgi2007shrec} has genus two as both hands are connected to the legs. The wrists are thinner than the other main shape structures so the arms are separated from the legs for the inner iso-surface at level $0.4$. In the bottom row, considering the teapot shape from Watertight dataset, we compute the curvature-sensitive distance field outside the shape volume. The level surfaces of the field are the boundaries of the {inflated} forms. The iso-surface at level $0.43$ has genus one since the upper joint of the handle is connected to the body.

%We solve for the distance field once for each shape to obtain the level surfaces simultaneously. We give the details of the field computation and extraction of the level surfaces in \S\ref{ssec:field}.

Once we have a collection of level surfaces for each shape, we determine the levels at which the iso-surfaces from both shapes are manifold meshes with the same genus number.  We compute the genus number of surfaces using the Euler formula after checking their manifoldness. Note that the level surfaces are always closed and connected meshes but they may include some non-manifold vertices or edges. Among the pairs of level surfaces, we consider the one with the smallest genus number at the smallest level for applying the 3D shape correspondence algorithm. The mapping computed between the selected pair of level surfaces  is transferred to the input shapes by computing their vertices closest to the matched points of the level surfaces.

\begin{figure} [t!]
\begin{center}
\includegraphics[width=\linewidth]{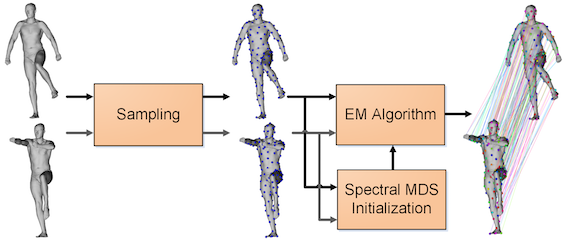}
\end{center}
   \caption{Block diagram of the EM algorithm \cite{ys2012EM}.}
\label{fig:em}
\end{figure}

{To produce the mapping between the pair of level surfaces extracted from our $v$ field \eqref{eq:energy_field}, we employ the isometric shape correspondence algorithm in \cite{ys2012EM} whose block diagram is given in Figure~\ref{fig:em}. Extracted  level surfaces have trustworthy geodesic distances. This is important for the isometric correspondence algorithm to perform well. The algorithm starts by defining an isometric distortion function that measures, for a given map, deviation from isometry, or equivalently, that quantifies the quality of a given map. The basic idea is then to search the space of all possible maps to minimize this isometric distortion. This minimization problem is cast as maximization of the likelihood function in a probabilistic setting that is solved via Expectation-Maximization (EM) algorithm.}

Given source and target meshes equipped with geodesic distances, the EM algorithm alternates between i) recomputing the expected value of a matrix that encodes the probability of source vertex $s_i$ being in correspondence with target vertex $t_j$ (E-step), and ii) estimating a mapping that maximizes the log-likelihood by using first bipartite perfect matching, and then a greedy optimization algorithm (M-step). The probability matrix, hence the EM algorithm, is initialized based on the Euclidean distances between the vertices embedded into spectral domain through classical Multi-Dimensional Scaling (MDS). It is then filled based on the current mapping produced by the M-step. As the M-step produces more reliable maps, the probability matrix, \emph{i.e.}, the E-step, becomes more accurate, which in turn leads to an even better M-step. This alternating optimization between E-step and M-step yields the optimal one-to-one mapping in the minimum-distortion sense.

The isometric distortion function that guides the optimization gives the difference between the pairwise geodesic distances between sampled points on the source shape, and the distance of their images under the map, aggregated over all pairs of sampled points using a standard norm. Specifically, 
\begin{equation}
D_{\mathrm{iso}}(\S) = \frac{1}{|\S|} \sum_{(s_i, t_j) \in \S} d_{\mathrm{iso}}(s_i, t_j)\label{eq:iso}
\end{equation}
where $\S$ denotes the set of correspondence pairs between $S$ and $T$, and
\begin{equation}
d_{\mathrm{iso}}(s_i, t_j) = \frac{1}{|\S|-1} \sum_{\substack{(s_l, t_m) \in \S\\(s_l, t_m)\neq(s_i, t_j)}} |g(s_i, s_l) - g(t_j, t_m)|
\label{eq:diso}
\end{equation}
where $g(.,.)$ is the geodesic distance between two base vertices, or more generally, between two points on a given surface. Hence, $d_{\mathrm{iso}}(s_i, t_j)$ is the contribution of the individual correspondence $(s_i, t_j)$ to the overall isometric distortion. Both $d_{\mathrm{iso}}$ and $D_{\mathrm{iso}}$ take values in the interval $[0,1]$ since the function $g$ is normalized with respect to the maximum geodesic distance over the surface. Note that the entries in the probability matrix are defined in terms of isometric distortion $d_{\mathrm{iso}}(s_i, t_j)$; namely the value $e^{-d_{\mathrm{iso}}(s_i,t_j)}$ is used as the probability of matching $s_i$ to $t_j$.

{Although the EM algorithm rests on the basic assumption that the shapes to be matched are perfectly isometric, the experiments conducted in \cite{ys2012EM} show that it performs well also on nearly isometric shapes, e.g., a male matching with a female. In the case of severe deviations from isometry, e.g., a cat and a giraffe, however,  the initially selected samples can be in very different configurations on the two surfaces so that unintuitive matchings can be generated as the output of the algorithm. We finally note that the method can handle input meshes with arbitrary genus.}

%-------------------------------------------------------------------------

\section{Experimental Evaluation}
\label{sec:evaluation}
%------------------------------------------------------------------------
\paragraph*{Datasets.}
We evaluate our method using four different datasets: Watertight dataset~\cite{giorgi2007shrec}, SHREC10 correspondence benchmark~\cite{bronstein2010shrecCorr}, SHREC11 robustness benchmark~\cite{bronstein2010shrec}, and a custom database TN-SCAPE which we constructed  by adding topological noise to five meshes from SCAPE dataset~\cite{anguelov2005scape}.

Original SCAPE dataset contains mesh models for a human body in different poses where there is a one-to-one correspondence between the models. TN-SCAPE is composed of five mesh models  from SCAPE and their five noisy versions which we constructed. 
The second data set, SHREC10 correspondence benchmark, includes three objects. Each object comes with a base shape model (null shape)  and five additional forms obtained from an isometric deformation of the null shape by adding topological noise of increasing strength. In our experiments, we match each null shape to the five models with topological noise.
The third data set, the retrieval training set of SHREC11 robustness benchmark,  contains twelve different shape models. For each model, there are one null shape, one isometric shape and five shapes with increasing degree of topological noise. We use a subset of these shapes for which the ground truth correspondence is available. 
For  the three out of four datasets (SHREC10, SHREC11, and TN-SCAPE), ground truth correspondence information is available.

\paragraph*{Evaluation Metric.}
For  three datasets for which the ground truth correspondence information is available,  we measure the performance of our correspondence method in terms of the deviation from the ground truth correspondence. Let $\S$ be the set of correspondence pairs between source and target shape. The average ground truth correspondence error is defined as
\begin{equation}
D_{\mathrm{grd}} = \frac{1}{|\S|} \sum_{(s_i, \, t_j) \in \S} {g(t_i, \, t_j)}
\label{eqn:Dgrd}
\end{equation}
where $(s_i, \, t_i)$ represents the ground truth correspondence and $g(t_i, \, t_j)$ is the geodesic distance between the vertices $t_i$ and $t_j$ on the target shape.

For the first data set, Watertight, ground truth correspondence information is not available; therefore, in evaluating the matching results for the shape pairs from this dataset, we present only visual matching results for the topologically different human and teapot shape pairs.

Since our focus is on the sparse correspondence between sample vertices of the input shapes, for better interpretation of errors, we consider the normalized average ground truth error ${\widetilde{D}_{\mathrm{grd}}}$ formulated as the average ground truth error $D_{\mathrm{grd}}$ divided by the sampling radius $r$. The sampling algorithm cannot guarantee the same sampling on the two shapes so ${\widetilde{D}_{\mathrm{grd}}} \le 1$ holds for the optimal mapping.

%-------------------------------------------------------------------------

%-------------------------------------------------------------------------
\section{Results and Discussion}
\label{sec:results}
%Our correspondence method handles topology noise by incorporating topologically similar interpretations of two shapes. 
\noindent We first demonstrate our approach using visual examples. Then, we present three groups of numerical evaluation tests using the three datasets for which the ground truth correspondences are available.  In all of our experiments, the dimension of the spectral domain is $6$. {In the experiments with SHREC11 robustness benchmark and TN-SCAPE dataset, we used $80$ sample vertices where the number of mesh vertices is $1.5$K and $12.5$K, respectively. In the experiments with SHREC10 correspondence benchmark, we used $280$ sample vertices where the number of mesh vertices ranges from $20$K to $50$K.}

\begin{figure*}[htb]
  \centering
  \begin{tabular}{ccc}
  \includegraphics[width=.23\linewidth]{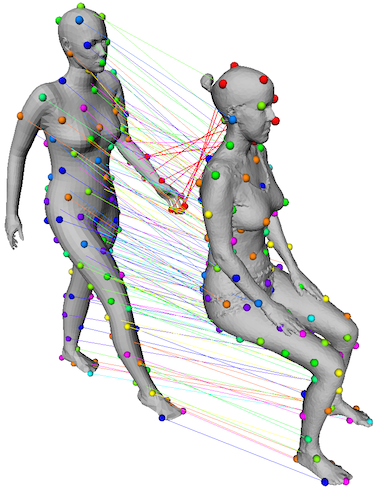} &
  \includegraphics[width=.23\linewidth]{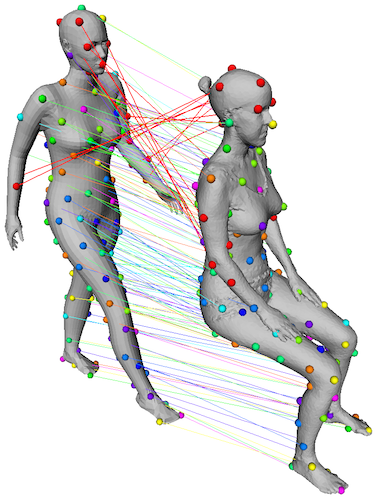} &
  \includegraphics[width=.46\linewidth]{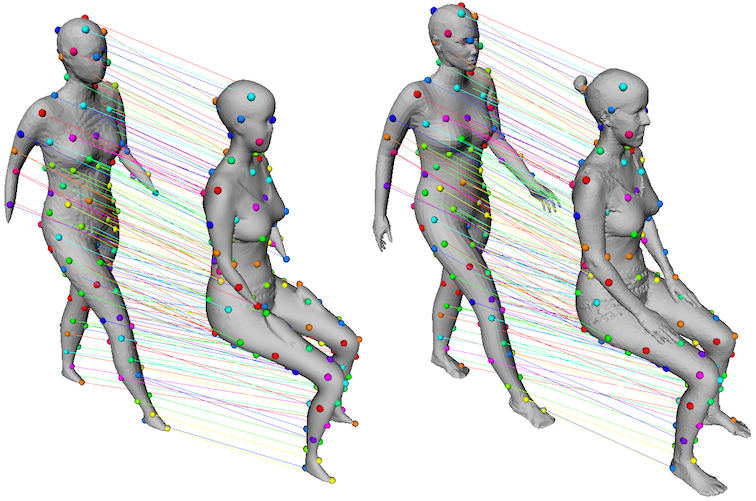}\\
  (a) & (b) & (c)
  \end{tabular}
  \caption{\label{fig:watertight_16_3}
  One-to-one mappings for a pair of human shapes from Watertight with different topology where both hands of the sitting woman are attached to the legs. (a) Correspondence between topologically different shapes using geodesic distance leads to errors where the right arm of the first shape is unmatched and the left arm is matched to the head of the sitting woman (mappings in red color). (b) Correspondence using biharmonic distance results in mapping the head of each shape to the arms of the other shape (mappings in red color). (c) Correspondence between inner iso-surfaces at level $0.4$ leads to a better result (left). Note that deflated forms of the shapes have the same topology as the arms of the sitting woman are separated from the legs. The mapping between the iso-surfaces is transferred to the input shapes (right).}
\end{figure*}
\begin{figure*}[htb]
  \centering
  \begin{tabular}{ccc}
  \includegraphics[width=.22\linewidth]{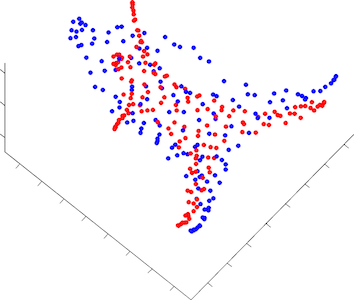} &
  \includegraphics[width=.22\linewidth]{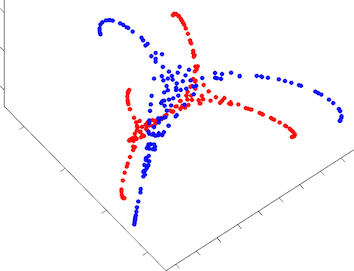} &
  \includegraphics[width=.22\linewidth]{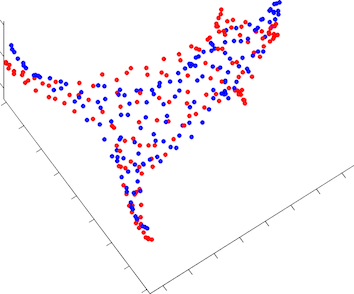} \\
  (a) & (b) & (c)
  \end{tabular}
  \caption{\label{fig:watertight_16_3_embed}
  Spectral embeddings and alignments that lead to the one-to-one mappings shown in Figure~\ref{fig:watertight_16_3}. Only the first $3$ dimensions are plotted. Red and blue points correspond to the standing and sitting woman, respectively. Sample vertices on the input shapes are embedded (a) using their pairwise geodesic distance (b) using their pairwise biharmonic distance. (c) Sample vertices on the inner iso-surfaces are embedded using their pairwise geodesic distance.}
\end{figure*}

Our approach employs the matching algorithm \cite{ys2012EM} on deflated or inflated forms of the pair of input models and then transfers the resulting mapping to the input models. Therefore, in order to evaluate  our contribution, we compare our results with two other mappings, one using geodesic distance and the other using  biharmonic distance~\cite{Lipman10}, both  obtained by applying the same correspondence algorithm (that we use) directly on the input models.

The biharmonic distance is insensitive to small topology changes~\cite{Lipman10}; thus, it is expected to work well. Note that replacing the topology-sensitive geodesic distance with a topologically robust one is a common approach for handling the topological noise~\cite{Bronstein10, Sharma10, Sharma11}. We also give an illustrative result for comparing our method with the one~\cite{Mateus08} that performs the best in topology noise category of SHREC10 correspondence benchmark.

In Figure~\ref{fig:watertight_16_3}, the resulting one-to-one mappings for a pair of human shapes from Watertight dataset are presented. The input models have different topologies as the arms of the sitting woman are merged with the legs. As shown in Figure~\ref{fig:watertight_16_3_embed}~(a)~and~(b), for both geodesic and biharmonic mapping, the connection between the arms and the legs of the sitting woman is reflected in the spectral domain (see the blue points representing the embedded sample vertices of the sitting woman). This connection affects both normalization and alignment of the embedded vertices. The initial alignment in Figure~\ref{fig:watertight_16_3_embed}~(a) leads to the local optimum in which the legs are correctly mapped but one arm of the first shape is unmatched and the other arm is matched to the head of the second shape, which in turn degrades the quality of the geodesic-based correspondence algorithm (see Figure~\ref{fig:watertight_16_3}~(a)). Similarly, the erroneous initial alignment in Figure~\ref{fig:watertight_16_3_embed}~(b) based on biharmonic distances yields the unsatisfactory mapping in Figure~\ref{fig:watertight_16_3}~(b) where the head of each shape is matched to the arms of the other shape. Note that the topological difference implies distortion of the shape isometries which leads to bad initialization and erroneous convergence of the correspondence algorithm. Our approach considers topologically similar representations of the input shapes and finds a mapping between the deflated forms of the input models computed as the inner iso-surfaces at level $0.4$. The topological similarity between the iso-surfaces enables a better initialization as in Figure~\ref{fig:watertight_16_3_embed}~(c) and yields the correct mapping in Figure~\ref{fig:watertight_16_3}~(c).

\begin{figure}[t!]
  \centering
  \begin{tabular}{cc}
  \includegraphics[width=.3\linewidth]{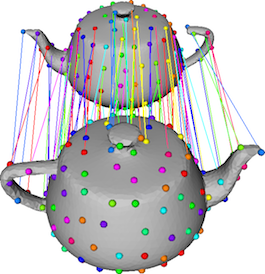} &
  \includegraphics[width=.6\linewidth]{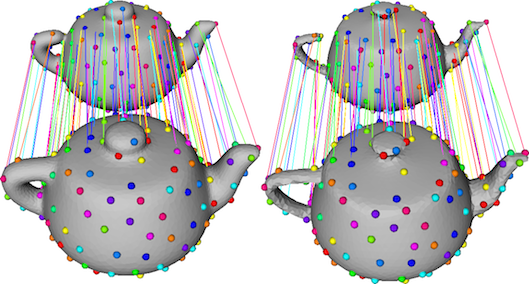} \\
  (a) & (b)\\
  \end{tabular}
  \caption{\label{fig:watertight_367_379}
  One-to-one mappings for a pair of teapot shapes from Watertight with different topology. The upper joint of the handle is cut in the farther shape. (a) Correspondence between topologically different shapes using geodesic distance leads to errors where the spouts are matched with the handles. (b) Correspondence between outer iso-surfaces at level $0.43$ leads to a better result as the handles and spouts are correctly matched (left). Note that inflated forms of the shapes have the same topology as open end of the farther shape handle joins the main body. The mapping between the iso-surfaces is transferred to the input shapes (right).
  }
\end{figure}

%%% Asli, bu iki figuru minipage icine koydum (sen bir bak uygun mu), siralari bozulmasin diye. \FloatBarrier calismiyor nedense ...
%\begin{figure*}[htb]
%\centering
%\begin{minipage}[b]{0.45\linewidth}
%  \centering
%  \begin{tabular}{cc}
%  \includegraphics[width=.32\linewidth]{figures/fig_watertight_367_379/watertight_367_379_orig.png} &
%  \includegraphics[width=.64\linewidth]{figures/fig_watertight_367_379/watertight_367_379_prop_all.png} \\
%  (a) & (b)\\
%  \end{tabular}
%  \caption{\label{fig:watertight_367_379}
%  One-to-one mappings for a pair of teapot shapes from Watertight with different topology. The upper joint of the handle is cut in the farther shape. (a) Correspondence between topologically different shapes using geodesic distance leads to errors where the spouts are matched with the handles. (b) Correspondence between outer iso-surfaces at level $0.43$ leads to a better result as the handles and spouts are correctly matched (left). Note that inflated forms of the shapes have the same topology as open end of the farther shape handle joins the main body. The mapping between the iso-surfaces is transferred to the input shapes (right).
%  }
%  \end{minipage}
%  \hskip 1cm
%  \begin{minipage}[b]{0.4\linewidth}
%    \includegraphics[width=1\linewidth]{figures/fig_shrec10_ex/shrec10_ex_topo5.png}
%  \caption{\label{fig:shrec10ex}
%  One-to-one mapping obtained by our method for two shapes from SHREC10 correspondence benchmark where the sitting man has topological noise of degree five.
%  }
%    \end{minipage}
%\end{figure*}

Our approach can also handle shapes with holes or breaks by considering their inflated forms. In Figure~\ref{fig:watertight_367_379}, we present the geodesic mapping and our proposed mapping for a pair of teapot shapes from Watertight dataset. The teapot models are topologically different as the upper joint of the handle is cut in the farther shape. The geodesic mapping shown in Figure~\ref{fig:watertight_367_379}~(a) erroneously matches the spouts with the handles. Our approach considers the inflated forms of the input shapes computed as the outer iso-surfaces at level $0.43$. Note that the inflated models have the same topology as the open end of the broken handle joins the main body and we obtain the correct mapping given in Figure~\ref{fig:watertight_367_379}~(b).

\begin{figure}[t]
  \centering
  \includegraphics[width=.75\linewidth]{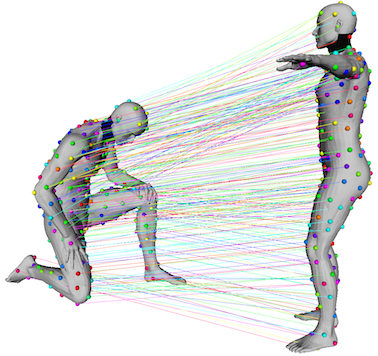}
  \caption{\label{fig:shrec10ex}
  One-to-one mapping obtained by our method for two shapes from SHREC10 correspondence benchmark where the sitting man has topological noise of degree five.
  }
\end{figure}

\begin{figure}[h!]
  \centering
  \includegraphics[width=.39\linewidth]{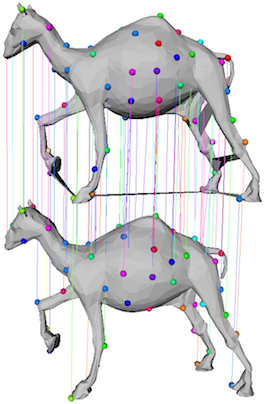}
  \includegraphics[width=.59\linewidth]{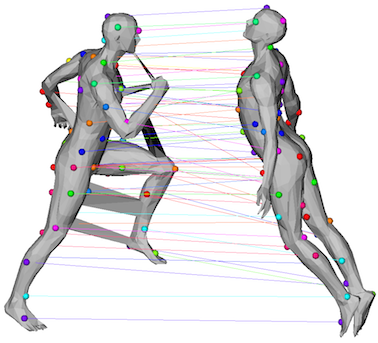}\\
  \includegraphics[width=.62\linewidth]{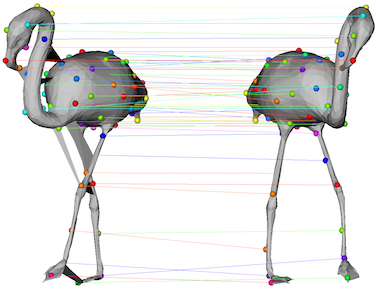}
  \caption{\label{fig:shrec11ex}
  One-to-one mappings obtained by our method for three pairs of shapes from SHREC11 robustness benchmark.
  }
\end{figure}

\begin{figure}[t]
  \centering
  \includegraphics[width=.84\linewidth]{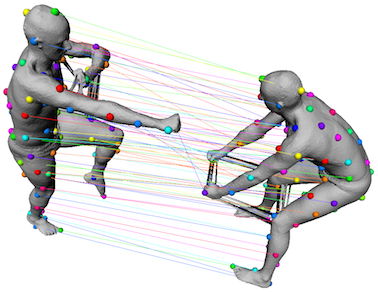}
  \caption{\label{fig:scape2ex}
  One-to-one mapping obtained by our method for a pair of shapes from TN-SCAPE dataset where both shapes have topology noise.
  }
\end{figure}

\begin{figure*}[t!]
  \centering
  \begin{tabular}{cc}
  \includegraphics[width=.39\linewidth]{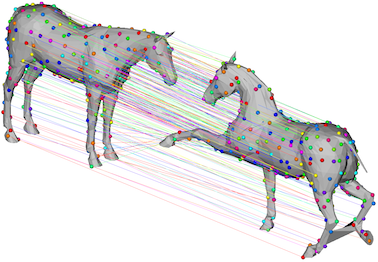} &
  \includegraphics[width=.39\linewidth]{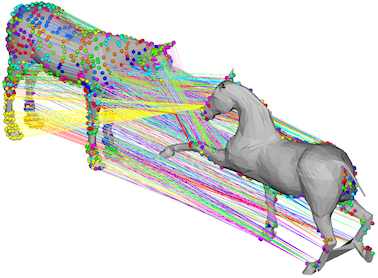}\\
  %(a) & (b)\\
  \end{tabular}
  \caption{\label{fig:mateus}
  A visual comparison of our approach with the method~\cite{Mateus08} that performs the best in the topology noise category of SHREC10 correspondence benchmark (Left) Our mapping result (Right) Dense mapping result obtained by the method~\cite{Mateus08}.
  }
\end{figure*}

In the first group of evaluation tests, we present the performance of our method in comparison with the geodesic-based and biharmonic-based mappings using SHREC10 correspondence benchmark. In the experiments, the null shape is mapped to each of the five shapes in the topological noise category. We examine how each method performs while the noise strength increases. In Table~\ref{table:shrec10res}, we present average of the normalized ground truth error ${\widetilde{D}_{\mathrm{grd}}}$ over the obtained results where the highest topology noise strength is different at each row. The geodesic-based mapping performs the best for the smallest noise degree but it immediately gets worse when the noise strength becomes greater than one. The biharmonic-based mapping diverges from being optimal when the noise strength is greater than three so it is more robust compared to the geodesic-based one. Our method is robust to the topological noise as all of the mappings are very close to the optimal and it performs the best for all of the experiments except the one with the smallest noise degree. We also expect that our method performs better than the related work~\cite{Bronstein10} that handles topology noise by employing the diffusion-based distance in the correspondence algorithm of~\cite{bronstein2006generalized}. This expectation is because of the fact that our method performs better than the correspondence algorithm~\cite{ys2012EM} running with biharmonic distances (see Table~\ref{table:shrec10res} biharmonic vs. proposed), and yet \cite{Bronstein10} employs a reportedly worse correspondence algorithm \cite{bronstein2006generalized} running with the same biharmonic distances. In Figure~\ref{fig:shrec10ex}, we show our mapping result for two human shapes from SHREC10 correspondence benchmark where the sitting man has topological noise of degree five. {The symmetric flip in Figure~\ref{fig:shrec10ex}, as well as in Figures \ref{fig:shrec11ex}, \ref{fig:scape2ex} and \ref{fig:mateus}, does not interfere with the topological noise robustness feature of our algorithm as one can always alleviate the symmetric flips by employing a denser sampling~\cite{Sahillioglu12b}.}
\begin{table}[t]
\centering
\begin{tabular}{|c|c|c|c|}
 \hline
 Noise strength & geodesic & biharmonic & proposed \\
 \hline
 $= 1$ & $\mathbf{0.86}$ & $1.67$ & $1.12$ \\
 \hline
 $\le 2$ & $2.15$ & $1.69$ & $\mathbf{1.13}$\\
 \hline
 $\le 3$ & $2.59$ & $1.70$ & $\mathbf{1.12}$\\
 \hline
 $\le 4$ & $3.58$ & $2.22$ & $\mathbf{1.12}$\\
 \hline
 $\le 5$ & $3.59$ & $2.53$ & $\mathbf{1.11}$\\
 \hline
\end{tabular}
\caption{\label{table:shrec10res}
Performance of our method in comparison with the geodesic-based and biharmonic-based mappings using the topology noise category of SHREC10 correspondence benchmark. The results represent average of ${\widetilde{D}_{\mathrm{grd}}}$ over the mappings. The highest topology noise strength is different at each row.
}
\end{table}

\begin{figure*}[htb]
  \centering
  \includegraphics[width=.75\linewidth]{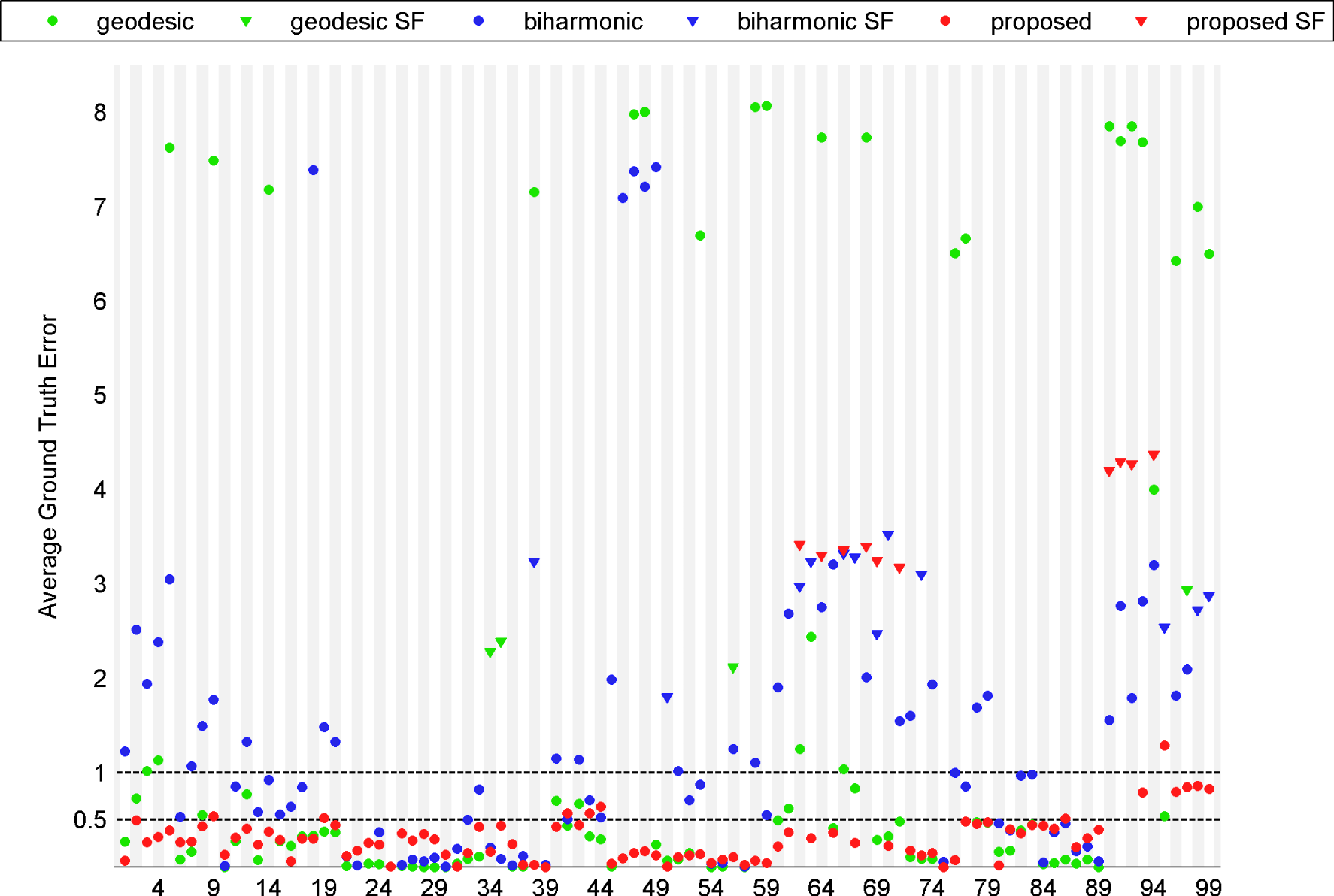}
  \caption{\label{fig:shrec11plot}
  Normalized average ground truth error ${\widetilde{D}_{\mathrm{grd}}}$ for one-to-one mappings between topologically different pairs of shapes from SHREC11 robustness benchmark. The errors in green and blue color are obtained using geodesic-based and biharmonic-based mapping, respectively. The errors in red color shows the performance of our approach. Some correspondence errors are high due to the symmetric flip (SF) problem indicated by triangles. Observe that our approach successfully handles topology noise by using topologically comparable forms of the input shapes.
  }
\end{figure*}
\begin{table}[t]
\centering
\begin{tabular}{|c|c|c|c|}
\hline
 & geodesic & biharmonic & proposed \\
 \hline
 {$\#$ of ${\widetilde{D}_{\mathrm{grd}}} \le 1$} & $69$ & $48$ & $88$ \\
 \hline
 $avg({\widetilde{D}_{\mathrm{grd}}})$ & $1.83$ & $1.37$ & $0.30$\\
 \hline
\end{tabular}
\caption{\label{table:shrec11res}
Summary of the results presented in Figure~\ref{fig:shrec11plot}. Performance of our proposed method in comparison with the geodesic and biharmonic mappings using SHREC11 robustness benchmark. In the first row, the number of optimal results for which ${\widetilde{D}_{\mathrm{grd}}}\le 1$ (over all of $99$ mappings) is given. In the second row, the average of ${\widetilde{D}_{\mathrm{grd}}}$ over all the mappings (excluding the results with symmetric flip) is given.}
\end{table}
In the next group of evaluation tests, we demonstrate the performance of our approach using the SHREC11 robustness benchmark. We use the shape models 0002, 0004, 0005, 0007, 0008, 0012 and 0014 for which the ground truth correspondence is available. For each model, we use the isometric shape and five shapes with topology noise. We also use the null shape from the models 0002 and 0007. In Figure~\ref{fig:shrec11plot}, we present the average ground truth error for the one-to-one mappings obtained using our proposed method, the geodesic-based and biharmonic-based mappings. The input pairs are the shapes from each model where at least one of them has topological distortion. As shown in Figure~\ref{fig:shrec11plot} and summarized in Table~\ref{table:shrec11res}, our approach successfully handles the topology noise as almost all of our mappings are optimal (${\widetilde{D}_{\mathrm{grd}}} \le 1$). Note that normalized average ground truth error is large for some results due to the symmetric flip problem. Excluding the mappings with symmetric flip, the average of the normalized average ground truth errors over all results, $avg({\widetilde{D}_{\mathrm{grd}}})$, is very small for our proposed method compared to the geodesic-based and biharmonic-based mappings (see Table~\ref{table:shrec11res}). Overall, the geodesic-based method either solves the correspondence problem or yields a result with an extremely high error. The biharmonic-based method has less number of optimal mappings but it generally solves some part of the correspondence problem and therefore decreases $avg({\widetilde{D}_{\mathrm{grd}}})$. Our method is robust to topological noise as it gives an optimal result for almost all of the mappings. In Figure~\ref{fig:shrec11ex}, we present one-to-one mappings obtained by our method for three pairs of shapes from SHREC11 robustness benchmark.

%\begin{figure}[htb]
%  \centering
%  \includegraphics[width=.9\linewidth]{figures/fig_shrec10_ex/shrec10_ex_topo5.png}
%  \caption{\label{fig:shrec10ex}
%  One-to-one mapping obtained by our method for two shapes from SHREC10 correspondence benchmark where the sitting man has topological noise of degree $5$.
%  }
%\end{figure}
%

\begin{figure}[t!]
  \centering
  \includegraphics[width=.17\linewidth]{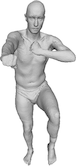}
  \includegraphics[width=.21\linewidth]{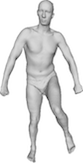}
  \includegraphics[width=.20\linewidth]{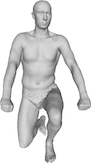}
  \includegraphics[width=.17\linewidth]{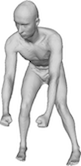}
  \includegraphics[width=.20\linewidth]{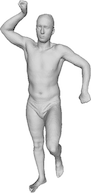}\\
  \includegraphics[width=.17\linewidth]{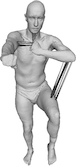}
  \includegraphics[width=.21\linewidth]{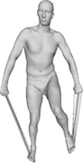}
  \includegraphics[width=.20\linewidth]{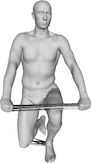}
  \includegraphics[width=.17\linewidth]{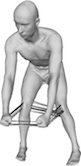}
  \includegraphics[width=.20\linewidth]{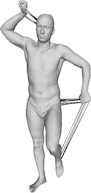}
  \caption{\label{fig:scapetopomodels}
  TN-SCAPE dataset that we construct by adding topology noise to five meshes from SCAPE. In the first row, the original shapes are shown and their counterparts with added topological noise are given in the second row.
  }
\end{figure}
\begin{figure}[t!]
  \centering
  \includegraphics[width=.72\linewidth]{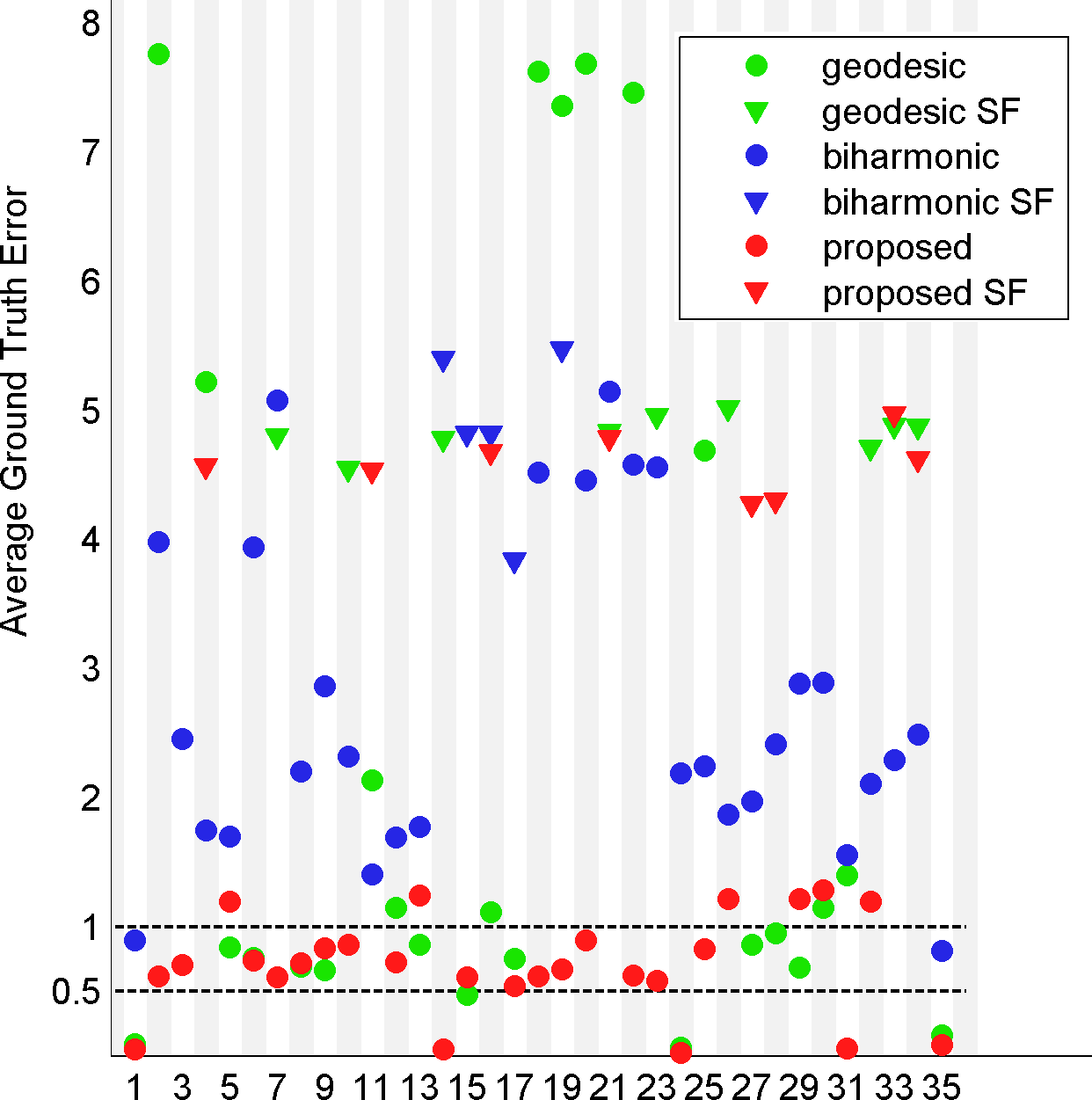}
  \caption{\label{fig:scapetopoplot}
  Normalized average ground truth error ${\widetilde{D}_{\mathrm{grd}}}$ for one-to-one mappings between topologically different pairs of shapes from TN-SCAPE dataset. The notations are the same as in Figure~\ref{fig:shrec11plot}. Observe that our approach successfully handles topology noise by using topologically comparable forms of the input shapes.
  }
\end{figure}
\begin{table}[t]
\centering
\begin{tabular}{|c|c|c|c|}
\hline
 & geodesic & biharmonic & proposed \\
 \hline
 {$\#$ of ${\widetilde{D}_{\mathrm{grd}}} \le 1$} & $14$ & $2$ & $21$ \\
 \hline
 $avg({\widetilde{D}_{\mathrm{grd}}})$ & $2.44$ & $2.70$ & $0.70$\\
 \hline
\end{tabular}
\caption{\label{table:scapetopores}
Summary of the results presented in Figure~\ref{fig:scapetopoplot}. Performance of our proposed method in comparison with the geodesic and biharmonic mapping using TN-SCAPE dataset. In the first row, the number of optimal results for which ${\widetilde{D}_{\mathrm{grd}}}\le 1$ (over all of $35$ mappings) is given. In the second row, the average of ${\widetilde{D}_{\mathrm{grd}}}$ over all the mappings (excluding the results with symmetric flip) is given.
}
\end{table}

In the final group of evaluation tests, we use TN-SCAPE dataset shown in Figure~\ref{fig:scapetopomodels} for further evaluation of our performance. The experimental settings are similar to the ones used in SHREC11 robustness benchmark. Again, we compare our mapping with the geodesic- and biharmonic-based mappings and we ensure that at least one shape in each input pair has topology noise. The experimental results are presented in Figure~\ref{fig:scapetopoplot}. In accordance with the previous results, our method successfully handles topology noise as most of the mappings are optimal (${\widetilde{D}_{\mathrm{grd}}} \le 1$) and few of them are very close to the optimal (${\widetilde{D}_{\mathrm{grd}}}$ is around $1$). Also, the average error measure $avg({\widetilde{D}_{\mathrm{grd}}})$ over all results, excluding the symmetric flips, is again very small for our proposed mapping compared to the geodesic- and biharmonic-based mappings (see Table~\ref{table:scapetopores}). In Figure~\ref{fig:scape2ex}, we show our proposed mapping for a pair of shapes from TN-SCAPE dataset where both shapes have topology noise.

Finally, we present a visual comparison of our approach with the method~\cite{Mateus08} that performs the best in the topology noise category of SHREC10 correspondence benchmark. We run the method~\cite{Mateus08} on a pair of horse shapes from SHREC11 robustness benchmark using its code available on the web. One of the horse shapes has the topological noise as its back legs are linked to each other. Figure~\ref{fig:mateus} shows that our approach successfully handles the topology noise whereas the method~\cite{Mateus08} fails to solve the correspondence problem under the given topology noise.

{We conduct our experiments on a $64$-bit workstation equipped with quad-core i7 processor (with clock frequency adjusted to 4.2GHz) and 32GB of RAM. We voxelize the mesh model of each shape so that the number of shape voxels is approximately $500$K. Computing the entire field (containing all deflations or inflations) over this voxel set takes, on the average,  $49$ seconds. Extracting a single level surface of the  field (a graded deflation or inflation) takes, on the average, $0.18$ seconds. }

%-------------------------------------------------------------------------
\vspace{-0.2cm}

\section{Conclusion}

We presented a simple yet effective approach for coping with topological noise when computing a correspondence between two shapes given in the form of surface meshes. Instead of constructing distance measures that are less sensitive to topological variations as compared to geodesic distance, we resort to graded deflation (inflation) of the object(s) to obtain meshes that are topologically comparable.  The deflated (inflated) shape boundaries at a variety of grades are simultaneously computed as the level surfaces of a field of which value at a point approximately codes the distance from the nearest boundary point as well the curvature of the respective level surface, hence, at the respective boundary location. Level surfaces obtained from the field can be analyzed by geodesic distances, which are known to be more accurate than diffusion based counterparts. As we use an approximate method for deflation (inflation), the computations are simple and fast. The step of computing a field and extracting all deflations or inflations for the purpose of equating topology adds an extra load which is on the order of a minute.  Considering that a computed field can be stored and later used in multiple matching tasks (not just for matching a single pair of shape), we find this cost negligible.

We  performed matching experiments using datasets for which ground truth mesh correspondences are available.  One of the test datasets is a small custom dataset (TN-SCAPE) prepared by us by adding topology noise to five of the model meshes of the SCAPE~\cite{anguelov2005scape} dataset. Experiments demonstrate that as the level of topological noise increase, our approach significantly outperforms the biharmonic distance.

\vspace{-0.3cm}

\section*{Acknowledgements} The work is funded by TUBITAK under grant 112E208 \url{http://user.ceng.metu.edu.tr/~asli/112E208.html}

%-------------------------------------------------------------------------

%\bibliographystyle{eg-alpha}
\bibliographystyle{eg-alpha-doi}

\vspace{-0.3cm}

%\bibliography{egbibsample_versionS2}

\newcommand{\etalchar}[1]{$^{#1}$}

\end{document}